\documentclass[12pt, preprint]{aastex}
\usepackage{natbib}

\shorttitle{Fermi Constraints and Cluster Radio Halos}

\shortauthors{}

\begin{document}

\title{Implications of {\em Fermi} Observations for Hadronic Models \\of Radio Halos in Clusters of Galaxies}

\author{Tesla E. Jeltema\altaffilmark{1} and Stefano Profumo\altaffilmark{2}}

\altaffiltext{1}{UCO/Lick Observatories, 1156 High St., Santa Cruz, CA 95064; tesla@ucolick.org}
\altaffiltext{2}{Department of Physics and Santa Cruz Institute for Particle Physics,  University of California, 1156 High St., Santa Cruz, CA 95064, USA; profumo@scipp.ucsc.edu}

\begin{abstract}
We analyze the impact of the {\em Fermi} non-detection of gamma-ray emission from clusters of galaxies on hadronic models for the origin of cluster radio halos. In hadronic models, the inelastic proton-proton collisions responsible for the production of the electron-positron population fueling the observed synchrotron radio emission yield a gamma-ray flux, from the decay of neutral pions, whose spectrum and normalization depend on the observed radio emissivity and on the cluster magnetic field. We thus infer lower limits on the average cluster magnetic field in hadronic models from the {\em Fermi} gamma-ray limits.  We also calculate the corresponding maximal energy density in cosmic rays and the minimal-guaranteed gamma-ray flux from hadronic radio-halo models. We find that the observationally most interesting cases correspond to clusters with large radio emissivities featuring soft spectra. Estimates of the central magnetic field values for those clusters are larger than, or close, to the largest magnetic field values inferred from Faraday rotation measures of clusters, placing tension on the hadronic origin of radio halos.  In most cases, however, we find that the {\em Fermi} data do not {\em per se} rule out hadronic models for cluster radio halos as the expected gamma-ray flux can be pushed below the \textit{Fermi} sensitivity for asymptotically large magnetic fields.  We also find that cosmic rays do not contribute significantly to the cluster energy budget for nearby radio halo clusters.
\end{abstract}

\keywords{acceleration of particles --- radiation mechanisms: non-thermal --- galaxies: clusters: general --- gamma rays: galaxies: clusters --- radio continuum: general}

\section{INTRODUCTION}

The observation of diffuse, Mpc scale radio synchrotron emission in clusters of galaxies indicates the presence of both relativistic particles and large-scale magnetic fields in the intra-cluster medium (ICM).  Morphologically, diffuse cluster radio emission is observed in two distinct varieties: radio relics and radio halos. Radio relics -- polarized, asymmetric structures typically at the outskirts of clusters -- may be associated to primary relativistic electron populations directly accelerated at merger or accretion shocks \citep[e.g.][]{1998A&A...332..395E, 2004rcfg.proc..335K}.  In contrast, radio halos are regular, unpolarized, and distributed similarly to the thermal X-ray emission from the ICM \citep[see e.g.][for recent reviews]{2008SSRv..134...93F, 2008LNP...740..143F}.  The large extent and regular appearance of radio halos is difficult to explain through a primary relativistic electron population as these particles are short lived, losing energy before they can diffuse through any significant distance in the cluster.  Therefore, relativistic electrons should trace where they are produced.  

A couple of models have been proposed to explain the existence of radio halos.  In the ``hadronic" model, secondary cosmic-ray electrons are generated in the collisions of relativistic protons with nuclei in the ICM \cite[e.g.][]{1980ApJ...239L..93D, 1999APh....12..169B}.  Cosmic-ray protons, like primary electrons, can be accelerated in merger and accretion shocks, by AGN, or by supernovae; unlike primary electrons, though, cosmic-ray protons loose energy less efficiently, diffusing for cosmological time-scales, thus effectively building up and fairly homogeneously populating the cluster halo.  A second possibility is that the relativistic electrons responsible for radio halo emission originate from lower energy electrons which are turbulently re-accelerated as a consequence of cluster mergers \cite[e.g.][]{2001MNRAS.320..365B, 2001ApJ...557..560P}.

It has been argued that the hadronic model has difficulty explaining the detailed properties of some radio halos \citep[see e.g.][for a review]{2004JKAS...37..493B} such as the radial extent of the radio emission \citep{2007A&A...465...41M, Donnert1}, the steepness of the radio spectra observed for some halos \citep{A2256, 2010arXiv1004.1515M}, and the steepening of the radio spectrum with increasing frequency or radius \citep{2004A&A...423..111F, Th03}(see however \cite{KL10}).  Other studies, however, have pointed out that hadronically generated cosmic-ray electrons emitting synchrotron radiation in the presence of a strong magnetic field ($B > 3 \mu$G) offer the most natural explanation of the strong correlation of radio halo power with cluster thermal X-ray luminosity and that merger induced enhancement and subsequent decay of the cluster magnetic field can explain the associations of radio halos to merging clusters \citep{KKW09, KL10} (see however \cite{2009A&A...507..661B}).  Models of cluster radio halos depend sensitively on the strength of the cluster magnetic field, on the cosmic-ray density, and on the distribution of both of these quantities throughout the cluster volume, none of which is robustly measured at present.  Some information on cluster magnetic fields can be gained through Faraday rotation measures of radio galaxies in or behind clusters, though the translation of rotation measures to magnetic field strength depends sensitively on the unknown spatial structure of cluster magnetic fields as well as on the gas density distribution \citep[see e.g.][and references therein]{2004IJMPD..13.1549G, 2002ARA&A..40..319C}.

In hadronic models of radio halos, the relativistic electrons responsible for the radio emission are generated through inelastic  collisions of high-energy cosmic-ray protons with nuclei in the intra-cluster gas.  These same collisions lead to gamma-ray emission through the production and subsequent decays of neutral pions.  Diffuse gamma-ray emission has yet to be detected from clusters \citep{egret, hesscoma, magicperseus, fermidm, fermiclus}, placing constraints on the possible energy density in cosmic-ray protons and therefore on the density of relativistic electrons that may be produced from these.  Recently, the \textit{Fermi}-LAT Gamma-ray Space Telescope has reported gamma-ray upper limits for a large sample of clusters \citep{fermiclus} which are roughly an order of magnitude more stringent than previous \textit{EGRET} limits \citep{egret}.  These limits imply that the energy density in cosmic-ray protons in nearby clusters is no more than a few percent of the thermal energy density, though the exact limits depend on the assumed cosmic-ray spectral and spatial distributions \citep{fermiclus}. In the present analysis we qualify the implications of the new gamma-ray observations for hadronic models (\cite{egret} point out that the \textit{EGRET} data ruled out the gamma-ray fluxes predicted in the hadronic models of \cite{ensslin97} and \cite{darshaviv95}, though these assume rather high cosmic ray densities.  \cite{egret} did not address the questions of the implications for the cluster magnetic fields or the minimal gamma-ray flux which we study here). Also, gamma-ray limits from clusters of galaxies have significant implications for particle dark matter searches \citep{fermidm,dmdecay}.

In this paper, we re-examine the hadronic scenario for the origin of radio halos in light of the tight upper limits placed by \textit{Fermi} on the possible gamma-ray emission from clusters of galaxies.  The gamma-ray non-detections effectively limit the possible density of secondary relativistic electrons and imply a minimum magnetic field strength if the observed radio halo fluxes are to be reproduced.  In \S2, we present our sample of radio halo clusters with \textit{Fermi} gamma-ray upper limits.  In \S3, we use the observed radio fluxes and radio spectra under the assumption that radio halos are hadronically produced to predict the minimum possible gamma-ray emission and to extract the minimal magnetic field strength required not to exceed the \textit{Fermi} gamma-ray limits.  In \S3.4, we compare the resulting magnetic field strengths required in the hadronic scenario to Faraday rotation estimates of cluster magnetic fields. Finally, we discuss our results and draw our conclusions in \S4. 

\section{CLUSTER SAMPLE}

We consider the sample of nine, nearby radio halo clusters which have published upper limits on their gamma-ray emission from the \textit{Fermi} Gamma-ray Space Telescope \citep{fermiclus}.  These clusters represent the best cases to test the hadronic production of radio halos.  We do not consider here cluster radio relics, since relics likely have a different origin (e.g.~primary electrons directly accelerated in merger shocks \citep[e.g.][]{1998A&A...332..395E}).  Most of the clusters in our sample host so-called giant radio halos with $\sim$ Mpc scale radio emission, although the halo in A2029 (and possibly in A2142) is classified as a mini-halo, i.e. a smaller diffuse radio source found at the center of cool core clusters \citep{M09, 2008LNP...740..143F}.

The basic radio properties of our cluster sample (the radio flux at 1.4 GHz and the power law spectral index) collected from the literature are listed in Table 1, which also indicates the upper limits on the gamma-ray flux in the 0.2-100 GeV band from \cite{fermiclus}.  In this work, we focus on whether the total radio flux and average radio spectrum of cluster radio halos can be reproduced in hadronic models given the \textit{Fermi} gamma-ray upper limits.  We will defer the consideration of the detailed spatial and spectral properties of radio halos, which may provide additional constraints, to future work.

\section{CONSTRAINTS ON HADRONIC MODELS}

\subsection{Radio Halo Modeling}
We approximate the primary proton cosmic-ray population with a power-law
\begin{equation}\label{eq:primary}
\frac{{\rm d}N_p}{{\rm d}E}=N_0(\vec r) \left(\frac{E}{\rm GeV}\right)^{-\alpha_p}
\end{equation}
with a low-energy cut-off set at $E\simeq0.3$ GeV, a value inferred from estimates of the minimal cosmic-ray energy provided by acceleration processes in clusters (e.g.~shocks and supernovae) that does not instantly thermalizes due to Coulomb and hadronic energy losses \citep[see e.g.][]{2008A&A...481...33J}. 

Electron-positron ($e^\pm$) production results from the decays of charged pions copiously created in inelastic scattering processes involving a primary cosmic-ray proton from the population in Eq.~(\ref{eq:primary}) and a target nucleus (essentially at rest) in the intra-cluster medium. The resulting secondary $e^\pm$ population features an approximate injection spectrum of the form \citep{1994A&A...286..983M, 1986A&A...157..223D}:
\begin{equation}\label{eq:injection}
\frac{{\rm d}N_e}{{\rm d}E}\propto \sigma_{pp}n_N(\vec r)N_0(\vec r)\left(\frac{E}{\rm GeV}\right)^{-\alpha_p},
\end{equation}
where $\sigma_{pp}\simeq 32$ mbarn indicates the inelastic proton-proton cross section at the energies of interest.

The population of high-energy $e^\pm$, which eventually will be responsible for the radio emission (as well as for inverse Compton emission potentially at energies as large as those detectable with {\em Fermi}), loses energy on short time scales corresponding to a spatial diffusion length much smaller than the size of a typical cluster \citep[see e.g.][]{1999ApJ...520..529S}. In other words, the effect of high-energy electron diffusion in clusters is always negligible \citep{berezinsky97,colablasi98}, and the steady-state solution for the equilibrium $e^\pm$ population results from accounting for energy losses alone, i.e.
\begin{equation}
n_e(\vec r,E_e)=\frac{1}{b(E_e)}\int_{E_e}^\infty\ {\rm d}E_e^\prime\ \frac{{\rm d}N_e}{{\rm d}E}(\vec r,E_e^\prime),
\end{equation}
where $b(E_e)$ indicates the energy loss term in the diffusion-loss equation \citep[see e.g.][]{2006A&A...455...21C}. We include in the function $b(E_e)$ all relevant energy loss processes, including inverse Compton scattering of CMB photons, synchrotron emission (assuming an average constant magnetic field in the region of interest), bremsstrahlung (using as an input the appropriate beta models for the gas density, from \cite{betamodel}) and Coulomb scattering. After having established the steady-state equilibrium $e^\pm$ solution, we calculate the synchrotron, inverse Compton and bremsstrahlung emission following the formalism outlined in \cite{2006A&A...455...21C} and recently reviewed in \cite{2010arXiv1001.4086P}.

The dominant gamma-ray emission process in the hadronic scenario is via the two-photon decay of neutral pions, which are produced with approximately the same multiplicity as their charged counterparts. To model the resulting gamma-ray spectrum we follow the analytic approximation to the results of \cite{1986A&A...157..223D} proposed in \cite{2004A&A...413...17P}. We verified that the relative normalization in the injection spectrum of gamma-rays from neutral pion decays versus electrons from charged pion decays follows the approximate relation, at $E\gg m_\pi$, \citep[see e.g.][]{2000A&A...362..151D}
\begin{equation}
\frac{{\rm d}N_\gamma}{{\rm d}E}(E)\simeq \frac{3}{16\alpha_p-8}\frac{{\rm d}N_e}{{\rm d}E}\left(\frac{E}{4}\right).
\end{equation}

\subsection{Limits on Cluster Magnetic Fields, Gamma-ray Emission, and Cosmic-Ray Energy Densities}

Fixing a value for the average magnetic field across a given radio halo, the observed radio flux yields the normalization to the primary cosmic-ray proton density. In turn, this produces a unique prediction (with the present assumptions on the morphology of cosmic-ray sources, and assuming a constant magnetic field intensity across the region of interest -- see the next section for the effect of a non-trivial magnetic field morphology) for the gamma-ray flux from neutral pion decay, as well as for the sub-dominant bremsstrahlung and inverse Compton emission processes from the secondary electron-positron population. Comparing this prediction to the {\em Fermi} limits on the gamma-ray flux from clusters exhibiting a radio halo, we can infer the minimal value of the cluster average magnetic field compatible with gamma-ray observations: since the radio emissivity is proportional to the square of the magnetic field intensity for a given electron density distribution, the smaller the magnetic field the larger the relative cosmic-ray proton primary population normalization, and hence the larger the gamma-ray flux from neutral pion decay.

We list in Table 2 our results for the constraints on hadronic models for cluster radio halos from {\em Fermi} observations. A crucial input parameter is the slope of the spectrum of the primary cosmic-ray protons, i.e. the parameter $\alpha_p$ in Eq.~(\ref{eq:primary}). We fix this parameter to match the slope of the radio emission at a given frequency, as listed in the fourth column of Table 1. For a power-law steady-state electron distribution with a spectral index $\alpha_e$, the radio emissivity has a corresponding spectral index \citep[see e.g.][]{1986rpa..book.....R} $$\alpha_R=(\alpha_e-1)/2.$$ For the present case, in the limit of high-energy, $b(E)\sim E^2$ and for the injection spectrum given in Eq.~(\ref{eq:injection}) one thus has $$n_e\sim\frac{1}{b(E)}\int_E (E^\prime)^{-\alpha_p}dE^\prime\sim E^{-(\alpha_p+1)},$$ which therefore implies $$\alpha_R\approx \alpha_p/2.$$ Of course, the radio spectrum depends on the assumed magnetic field as well as, potentially, on other energy loss terms contributing in $b(E)$ which are not quadratic in energy, such as bremsstrahlung, so the above is an approximation. Nonetheless, this is a good guideline to guess the input $\alpha_p$ for which one can obtain the correct radio spectrum, and we then iterate considering the full energy loss equation to find $\alpha_p$. Comparing the values of $\alpha_R$ from Table 1 and of $\alpha_p$ from Table 2, we do find a good confirmation of the theoretical approximation.

\cite{fermiclus} report \textit{Fermi} gamma-ray upper limits for a large sample of clusters in four energy bands in their Table 2.  Having fixed the appropriate cosmic-ray spectrum, we calculate the minimal average magnetic field such that the predicted gamma-ray flux does not exceed the limits quoted in any one of the four \textit{Fermi} bands. The resulting minimum average magnetic fields and the {\em Fermi} bands giving the most stringent constraints for each individual cluster are listed in the third and fourth columns of our Table 2, respectively. Notice that the values of the magnetic fields we quote here are ``average'' in the sense that we assume a single magnetic field value for the radio halo.

If cluster radio halos originate from hadronic cosmic-ray processes, the larger the average cluster magnetic field, the smaller the predicted gamma-ray flux. The latter, however, is bounded from below: we provide here a calculation of this minimal gamma-ray flux expected in the context of hadronic radio halo models. In the limit in which the magnetic field intensity is much larger than the equivalent magnetic field energy density in the CMB, $B^2\gg B_{\rm CMB}^2$, where $B_{\rm CMB}=3.2(1+z)^2 \mu$G, synchrotron dominates the energy losses of high-energy electrons, and $n_e\sim B^{-2}$. Since the radio emissivity is proportional to the magnetic field intensity squared and directly proportional to $n_e$, in the limit $B^2\gg B_{\rm CMB}^2$ the radio emissivity is essentially independent of the magnetic field intensity. Correspondingly, in this limit one gets the smallest possible value for the primary cosmic-ray proton density, and hence the minimal $\gamma$-ray flux in the context of the hadronic scenario for cluster radio halos. The fifth column of Table 2 indicates how far the current {\em Fermi} limits are from this minimal (``{\em guaranteed}'') flux of $\gamma$ rays, for the best {\em Fermi} energy band (indicated in the column to the left). We find that the current {\em Fermi} sensitivity is anywhere between 0.1\% and almost 10\% of this guaranteed $\gamma$-ray flux in the hadronic scenario, where the least promising clusters have the hardest cosmic-ray proton spectra ($\alpha_p\lesssim 3$) while the most promising clusters have the softest spectra ($\alpha_p\gtrsim3$). This is easily understood, since soft cosmic-ray spectra produce copious soft neutral pions that dump $\gamma$-rays within the low-energy range accessible to Fermi, while they hardly produce any electrons-positrons energetic enough to radiate synchrotron radio emission at the required level. This implies that to achieve the measured radio flux the normalization of the cosmic-ray primaries must be tuned to large values, yielding, in turn, very large $\gamma$-ray fluxes. We also indicate, in the sixth column of Table 2, the integrated minimum $\gamma$-ray flux for the hadronic model across the entire {\em Fermi} energy range (0.2-100 GeV), to be compared with the sixth column of Table 1 (showing, in the same energy range and with the same units, the current limits).

The last column of Table 2 shows our results for the maximal value of the ratio $\epsilon_{CR}/\epsilon_{th}$ of the cosmic-ray proton (kinetic) energy density $\epsilon_{CR}$ versus the cluster thermal energy density $$\epsilon_{th}=\frac{3}{2}\ d_e\  n_{e,th}(r)\  kT$$ compatible with the \textit{Fermi} limits.  Here, $n_{e,th}$ indicates the number density of thermalized electrons, $$d_e=1+\frac{1-\frac{3}{4}X_{\rm He}}{1-\frac{1}{2}X_{\rm He}}$$ is the number of particles per electron in the ICM, and we set the ${}^4$He mass fraction to its primordial value $X_{\rm He}\simeq0.24$. Finally, we assume that the ICM temperature $T$ is constant with radius.  To calculate the cosmic-ray energy density, we use the cosmic-ray primary spectrum outlined above (featuring a cut-off at 0.3 GeV for all clusters), with $\alpha_p$ set to the values indicated in the second column of Table 2.  We assume that the spatial distribution of the primary cosmic-ray protons follows the intra-cluster thermal gas density (i.e.: $N_0(\vec r)\propto n_{e,th}(\vec r)$), which we model using an isothermal beta-model.  We employ the temperature values and beta-model parameters listed in \cite{betamodel}. Notice that with our assumption of cosmic-ray proton sources following the ICM gas density, and of negligible cosmic-ray diffusion, the ratio $\epsilon_{CR}/\epsilon_{th}$ is a constant in space. We also note that the figures we quote in the last column of Table 2 depend on our assumption that the cosmic ray proton density scales like the thermal gas density. Relaxing this assumption and allowing for a relative bias in the two density distributions might produce a scatter by a factor of a few in the constraints on the relative cosmic ray energy density, see e.g. \cite{2009PhRvD..80b3005J}.

We find upper limits on the ratio $\epsilon_{CR}/\epsilon_{th}$ ranging between 0.01 to over 1.0 for the radio halo clusters in our sample.  For the nearest radio halo clusters, cosmic rays contribute no more than 1-5\% of the thermal energy density \citep[see also][]{fermiclus}.  The values listed in column 7 of Table 2 are ``maximal'' in that they refer to the minimal value of the cluster magnetic field inferred from the radio emission: increasing the magnetic field suppresses the cosmic-ray normalization while not affecting substantially the associated spectral shape. This is illustrated visually, for the case of three clusters (Coma, A1914 and A2163) in Fig.~1, where we plot $\epsilon_{CR}/\epsilon_{th}$ as a function of the average cluster magnetic field. We indicate the impact of the \textit{Fermi} limits, which constrain the magnetic fields to be {\em above} certain values, and therefore the energy density ratio to be {\em below} certain corresponding levels. The figure illustrates in the hadronic model clusters with a soft cosmic-ray spectrum (A1914) feature very large relative cosmic-ray energy densities, which, for reasonable magnetic field values, are typically more than 10\% of the thermal energy density \citep[see also][]{A2256, 2010arXiv1004.1515M}. The opposite is true for clusters with hard cosmic-ray spectra. Notice that the relative normalization of the curves depends sensitively on the cluster radio emissivity, temperature and gas density profile.

It should be noted that radio halos are often observed to be associated to merging clusters \citep[e.g.][]{2010ApJ...721L..82C}, and several of the clusters in our sample exhibit non-trivial spatial variations in the ICM density and temperature.  In the hadronic model, gamma-ray emission and relativistic electrons result from the same proton-proton collisions, so the calculation of the minimum gamma-ray emission does not depend assumptions of the ICM distribution or temperature nor does the derivation of the minimum magnetic field.  We have made the simplifying assumption of a constant average magnetic field (note the minimum gamma-ray emission does not depend on the magnetic field), which is likely not the case, but without a resolved gamma-ray or non-thermal X-ray detection, spatial variation in the magnetic field and cosmic ray electron population cannot be separated; as clusters are more or less point sources for \textit{Fermi}, the gamma-ray limits give us only a single average constraint on the cosmic ray population.  Below we consider a couple of simple magnetic field radial profiles.  Variations in the ICM density and temperature will effect the estimate of the cosmic ray energy density and thermal energy density (though here are average temperature should be reasonably accurate).  Under the assumption made above that the cosmic ray density traces the thermal gas density, the effect is minimal, but as already mentioned biases in the location of the cosmic ray population relative to the thermal gas will effect the constraints (at the level of a factor of a few for cosmic ray distributions motivated by simulations \citep[e.g.][]{2008MNRAS.385.1211P, 2009PhRvD..80b3005J}).

\subsection{Estimates for the Central Magnetic Field}

The results we quote above refer to the simplified scenario where the cluster magnetic field is constant throughout the cluster. The limits on the magnetic field quoted in Table 2 obviously depend on this assumption.  In this section, we consider a more realistic radially varying magnetic field and, in particular, we extract estimates for the central magnetic field value.  In the next section, we then compare our minimum required magnetic fields in the hadronic model to Faraday rotation measurement determinations of cluster magnetic fields and ask whether the needed magnetic fields actually exist in clusters. 

Both MHD simulations and Faraday rotation measures, for the few clusters with rotation measures at multiple distances, imply that cluster magnetic fields scale with the ICM gas density and decrease with radius \citep{2001A&A...378..777D, 2004IJMPD..13.1549G, Donnert2, comaRM}.  Denoting with $F(r)$ the spatial distribution of the thermal cluster gas as given by a beta-model with core radius $r_c$, i.e. $$ \frac{n_N(r)}{n_N(0)}\simeq F(r)=\left(1+\left(\frac{r}{r_c}\right)^{2}\right)^{-\frac{3}{2}\beta}$$ and assuming that the magnetic field scales with the gas density, we employ a phenomenological radial dependence of the magnetic field strength -- to be motivated below -- of the type: $$B(r)=B_c [F(r)]^\eta.$$ In this case, the central value of the magnetic field $B_c$ can be expressed as a function of the average magnetic field $B_0$ (which we constrain in the previous section) as:
\begin{equation}
B^2_c=B_0^2\frac{\int_0^{r_{h}}r^2[F(r)]^2{\rm d}r}{\int_0^{r_{h}}r^2[F(r)]^{2+2\eta}{\rm d}r},
\end{equation}
where $r_h$ indicates the radial size of the radio halo. It is straightforward to generalize the equation above to any magnetic field spatial distribution.

We quote the constraints on the central magnetic field for two cases, $\eta=0.5$ and $\eta=1$, for the radio halo clusters under consideration here in the last two columns of Table 3. These values for $\eta$ effectively bound the values found in simulations and Faraday rotation measure studies. For example, $\eta=0.5$ provides a good match to the radial dependence of Faraday rotation measures observed for the Coma cluster in the recent study of \cite{comaRM}, while simulations find a value of $\eta$ close to one \citep{2001A&A...378..777D, Donnert2}, and $\eta=0.9$ is a good match to the rotation measures observed for the cluster A119 \citep{2001A&A...378..777D}.  A magnetic field frozen in to the plasma would give instead an intermediate value, $\eta=2/3$.  For reference, we note in the second and third columns of Table 3 the radio halo size we consider and the minimal average magnetic field, $B_0$, determined from the \textit{Fermi} data in the analysis in the previous section.

\subsection{Comparison to Faraday Rotation Measures}

The only cluster in our sample whose magnetic field has been studied in detail using Faraday rotation measures is A1656, also known as the Coma cluster \citep{comaRM, 1990ApJ...355...29K, 1995A&A...302..680F}.  The recent study of \cite{comaRM} analyzed polarization data for seven radio sources within the Coma cluster at varying radii to derive the radial profile of the magnetic field strength.  They find a best fit central magnetic field value of $B_c = 4.7 \mu$G for a radial dependence compared to the thermal gas of $\eta = 0.5$, consistent with but close to the value of $3.9 \mu$G for the same $\eta$ that we infer for the minimum central field in the hadronic scenario.  Similarly, \cite{comaRM} find an average magnetic field over the cluster volume of $\sim 2 \mu$G close to our minimum derived value of $B_0 = 1.7 \mu$G, and for $\eta=1$ our central magnetic field of $7.4 \mu$G is only marginally consistent with the rotation measure data ($\sim 1$\% confidence level).  Therefore, while the hadronic generation of the radio halo in the Coma cluster is still consistent with Faraday rotation measure estimates of its magnetic field, any improvement in the gamma-ray upper limits will make the needed magnetic field in hadronic models too high compared to Faraday rotation measure estimates.

Among our cluster sample, A1914 which hosts a very steep spectrum radio halo, notably exhibits a very high required magnetic field if the radio halo is of hadronic origin.  For this cluster, we derive a minimum average magnetic field of $7.4 \mu$G and a minimum central magnetic field in the range $21 - 40 \mu$G.  While Faraday rotation measure estimates of the magnetic field are not available for this cluster, these values are much higher than the typical few $\mu$G magnetic fields found in rotation measure studies of non-cooling flow clusters like A1914 (and of clusters hosting giant radio halos in general which tend to be mergers); Faraday rotation studies only imply such high values of the magnetic field for observations of radio galaxies in the centers of cool core clusters (in some cases hosting radio mini-halos) and not over such large scales \citep[e.g.][]{2008SSRv..134...93F, 2002ARA&A..40..319C, 2004JKAS...37..337C}.  The requirement of such a large average magnetic field in A1914, therefore, sheds doubts on a hadronic origin for this radio halo.  In addition to Coma and A1914, the two clusters A2256 and A2319 also have minimum required magnetic fields of a few $\mu$G which, while consistent with the typical magnetic fields derived from Faraday rotation measures, will similarly soon be in tension with these if clusters continue to be undetected with deeper {\em Fermi} gamma-ray observations, or perhaps with new Faraday rotation measure data.  A2319, for example, has an early Faraday rotation measure estimate of its magnetic field strength of $2 \mu$G \citep{1987ApL....25..181V}.

We warn the reader that a few assumptions (including the functional dependence of the magnetic field power spectrum and the magnetic field radial distribution) come into play when translating Faraday rotation measures into de-projected magnetic field values, see e.g. \cite{2004A&A...424..429M} and \cite{comaRM}. Finally, we note that equipartition estimates of the average magnetic fields in radio halo clusters are typically much smaller than those  derived from Faraday rotation measures, being in the range of $0.1-1 \mu$G \citep{2008SSRv..134...93F}, and are smaller than the minimum values derived here for many of the clusters in our sample.

\section{DISCUSSION}

In this paper, we analyze how early {\em Fermi} constraints on the gamma-ray flux from clusters of galaxies impact hadronic models for clusters radio halos. We consider all clusters with both a radio-halo detection and for which {\em Fermi} gamma-ray upper limits are available. We tune the cosmic-ray proton spectrum responsible, in hadronic models, for the subsequent production of the non-thermal electron-positron population that fuels the synchrotron emission observed at radio frequencies, to match the slope and intensity of the detected radio emission.   We then calculate the smallest possible value for the average magnetic field in the radio halo region compatible with both the total observed radio halo flux and with the {\em Fermi} limits on the gamma-ray emission which would inevitably result from the same inelastic proton-proton collisions producing the electron-positron population. We note that the resulting magnetic field constraints are especially interesting for those radio halos with a soft radio spectra. For clusters such as A1914 or A2256, we find that in order to avoid over-producing gamma rays at {\em Fermi} energies, average magnetic fields at least on the order of several micro-Gauss are needed.  Using two phenomenological models for the magnetic field radial dependence, the required central magnetic fields are a few $\mu$G for several clusters and over 20 $\mu$G for A1914.  We find that in a few cases the implied magnetic fields in hadronic models are in excess of, or close to, the largest cluster magnetic field values obtained with Faraday rotation measures, placing tension on the hadronic origin of radio halos. The large values of the average magnetic fields we infer also imply a large non-thermal pressure, well beyond equipartition, that would have dramatic implications for the assumption of hydrostatic equilibrium and for cluster mass estimates and their use for cosmological studies. Further information may be gained by considering spatial variation in the observed radio flux and spectrum for specific selected clusters with the required wealth of radio data, and we defer this question to a forthcoming study (Jeltema et al, in prep.).

We also estimate the ratio of the energy density in cosmic rays over the cluster thermal energy density -- a quantity that is {\em maximal} for the {\em minimal} value of the magnetic field we infer. We find that this ratio varies between a percent to values of order unity, with larger values associated to both very hard (cosmic-ray spectral index $\alpha_p\approx 2$) or very soft ($\alpha_p\gtrsim 3$) cosmic-ray -- and hence radio -- spectra.  For at least a few nearby radio halo clusters it is clear that cosmic rays do not contribute significantly to the cluster energy budget \citep[see also][]{fermiclus}.  The minimal gamma-ray emission expected in the context of hadronic radio-halo models, corresponding to asymptotically large values of the cluster average magnetic field, is also derived. This minimum emission is typically between one to three orders of magnitude below the current {\em Fermi} sensitivity. The most promising cases are those corresponding to clusters with soft radio spectra, for which the needed improvement in sensitivity to detect a gamma-ray emission associated to a hadronic radio halo for \textit{any} magnetic field is on the order of 10. 

In summary, we showed that while hadronic models for radio halos are not ruled out by {\em Fermi} gamma-ray limits on galaxy clusters, the required cluster magnetic fields are in some cases high and possibly in tension with other observations.  This result reenforces how the high-energy end of the electromagnetic spectrum provides an important and crucial tool to understand the nature of non-thermal phenomena in the largest bound structures in the universe.

\acknowledgments
This work is partly supported by NASA grants NNX09AT96G and NNX09AT83G. SP also acknowledges support from the National Science Foundation, award PHY-0757911-001, and from an Outstanding Junior Investigator Award from the Department of Energy, DE-FG02-04ER41286.



\begin{deluxetable}{lccccc}
\tablecaption{ Cluster Data }
\tablewidth{0pt}
\tablecolumns{6}
\tablehead{
\colhead{Cluster} & \colhead{z} & \colhead{1.4 GHz Flux} & \colhead{$\alpha_R$} & \colhead{Ref.} & \colhead{$\gamma$-ray Upper Limit} \\
\colhead{} & \colhead{} & \colhead{Density (Jy)} & \colhead{} & \colhead{} & \colhead{($10^{-9}$ph cm$^{-2}$ s$^{-1}$, 0.2-100 GeV)}
}
\startdata
A1656 &0.023 &0.64 &1.35 &1 &4.58 \\
A1914 &0.171 &0.064 &1.88 &2, 3 &1.19 \\
A2029 &0.076 &0.022 &1.35 &4, 5 &3.28 \\
A2142 &0.090 &0.0183 &1.5 &6 &2.82 \\
A2163 &0.203 &0.155 &1.18 &7, 8 &5.51 \\
A2256 &0.060 &0.103 &1.61 &9 &1.96 \\
A2319 &0.056 &0.212 &1.28 &4, 10 &0.75 \\
A2744 &0.308 &0.0571 &1.0 &11, 12 &2.49 \\
1E 0657-56 &0.296 &0.078 &1.3 &13 &2.75 \\
\enddata
\tablecomments{Gamma-ray upper limits are taken from \cite{fermiclus}.  Radio references: (1) \cite{Th03}, (2) \cite{2003A&A...400..465B}, (3) \cite{G09}, (4) \cite{M09}, (5) \cite{1983PASAu...5..114S}, (6) \cite{2000astro.ph..8342G}, (7) \cite{2004A&A...423..111F}, (8) \cite{2001A&A...373..106F}, (9) \cite{2008A&A...489...69B}, (10) \cite{2001ApJ...548..639K}, (11) \cite{2001A&A...376..803G}, (12) \cite{2007A&A...467..943O}, (13) \cite{2000ApJ...544..686L}.}
\end{deluxetable}

\clearpage

\begin{deluxetable}{lcccccc}
\label{tab:hadro}
\tablecaption{ Hadronic Model Results }
\tablewidth{0pt}
\tablecolumns{6}
\tablehead{
\colhead{Cluster} & \colhead{$\alpha_p$} & \colhead{$B_{0,min}$}  & \colhead{\textit{Fermi} Band} & \colhead{Ratio of $F_{\gamma, min}$} & \colhead{Minimal $\gamma$-ray Flux}  & \colhead{Maximum}\\
\colhead{} & \colhead{} & \colhead{($\mu$G)}  & \colhead{(GeV)} & \colhead{to \textit{Fermi} up. lim.} & \colhead{($10^{-9}$ph cm$^{-2}$ s$^{-1}$)} & \colhead{$\epsilon_{CR}/\epsilon_{th}$}
}
\startdata
A1656 &2.70 &1.70  &1-10 &0.040 & 0.186 &0.018 \\
A1914 &3.76 &7.40 &0.2-1&0.077 & 0.181 &0.684 \\
A2029 &2.70 &0.37  &0.2-1 &0.001& 0.006 &0.052 \\
A2142 &3.01 &0.65 &0.2-1&0.002 & 0.010 &0.091 \\
A2163 &2.05 &0.37  &0.2-1&0.004 & 0.011 &0.131 \\
A2256 &3.25 &2.70 &0.2-1&0.037 & 0.099 &0.138 \\
A2319 &2.57 &1.40 &0.2-1 &0.040 & 0.046 &0.010 \\
A2744 &2.01 &0.26 &1-10 &0.004 & 0.004 &1.396 \\
1E 0657-56 & 2.60 & 0.77  & 1-10 & 0.005 & 0.018 & 0.508 \\
\enddata
\tablecomments{Column 2 lists the cosmic-ray proton injection spectral index implied by the radio spectrum for a magnetic field strength of $B=1 \mu$G.  Column 3 indicates the minimum average magnetic field required to produce the observed radio halo flux with secondary electrons without overproducing the gamma-ray flux compared to the \textit{Fermi} limits.  Column 5 gives the ratio of the minimum gamma-ray flux for any magnetic field value, corresponding to very large magnetic field values, in the hadronic model compared to the \textit{Fermi} limits. Columns 3 and 5 are calculated for the most constraining \textit{Fermi} energy band from \cite{fermiclus}, which is in turn noted in Column 4.  Column 6, instead, gives the minimal integrated gamma-ray flux between 0.2 and 100 GeV predicted in the hadronic scenario. The final column lists the maximum allowed ratio of the energy density in cosmic rays to the cluster thermal energy density implied by the \textit{Fermi} limits given the proton spectral index implied by the radio spectrum and assuming that the cosmic-ray density traces the thermal gas density.}
\end{deluxetable}

\clearpage

\begin{figure}
\plotone{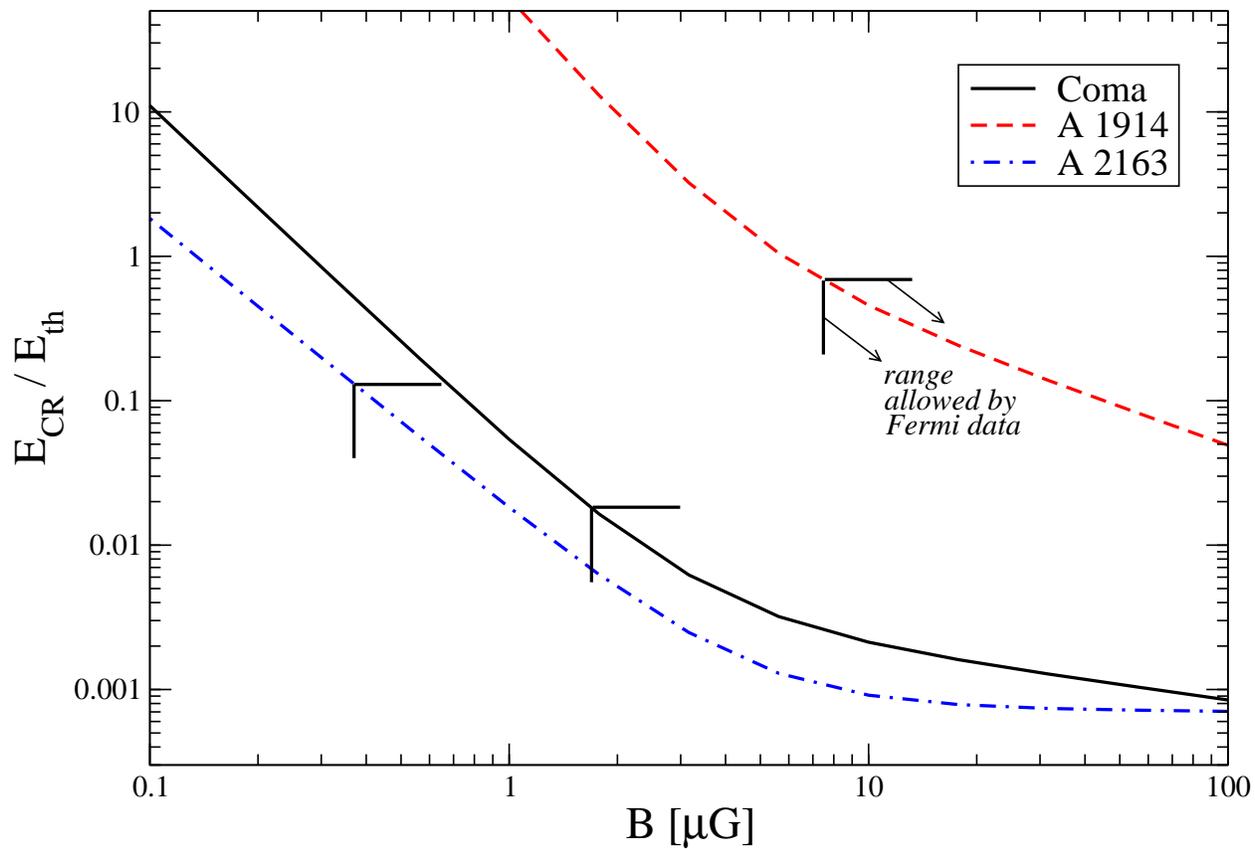}
\caption{Ratio of the energy density in cosmic rays to the thermal energy density as a function of the average magnetic field, $B_0$, for Coma (black solid line), A1914 (red dashed line) and A2163 (dot-dashed blue line).  Also indicated on each curve are the minimum average magnetic field and corresponding maximum allowed energy density ratio needed to produce the radio halo flux without overproducing the \textit{Fermi} gamma-ray upper limits.}
\end{figure}

\clearpage

\begin{deluxetable}{lcccc}
\label{tab:bfields}
\tablecaption{ Magnetic Field Model Results }
\tablewidth{0pt}
\tablecolumns{5}
\tablehead{
\colhead{Cluster} & \colhead{$r_h$} &  \colhead{$B_{min}$}  & \colhead{$B_c$, $\eta=1.0$} & \colhead{$B_c$, $\eta=0.5$} \\
\colhead{} & \colhead{(Mpc)} & \colhead{($\mu$G)}  & \colhead{($\mu$G)}  & \colhead{($\mu$G)}  
}
\startdata
A1656 & 0.83 &1.70  & 7.24 & 3.91\\
A1914 & 1.04 &7.40 & 40.1 & 21.1 \\
A2029 & 0.25 &0.37  & 1.80 & 0.93\\
A2142 & 0.19 &0.65 & 1.51& 1.01\\
A2163 & 2.28 &0.37  & 1.92& 1.03\\
A2256 & 0.81 &2.70 & 8.48& 5.12\\
A2319 & 1.02 &1.40 & 7.62& 3.82\\
A2744 & 1.89 &0.26 & 1.15& 0.65\\
1E 0657-56 & 2.10 & 0.77  & 5.73& 2.78\\
\enddata
\tablecomments{Column 2 indicates the size of the radio halo we employ \citep{G09, M09, 2000astro.ph..8342G}. Column 3 lists the minimum average magnetic field required to produce the observed radio halo flux with secondary electrons without overproducing the gamma-ray flux compared to the \textit{Fermi} limits. Column 4 and 5 list the predictions for the minimal value of the cluster central magnetic field, for $\eta=1.0$ and $\eta=0.5$, where $\eta$ indicates the bias in the magnetic field radial distribution with respect to the the cluster gas density distribution modeled as a beta-model.}
\end{deluxetable}

\end{document}